\documentclass{aa}

\usepackage[varg]{txfonts}
\usepackage{siunitx}
\usepackage[version=4]{mhchem}
\usepackage{multirow}
\usepackage{graphicx}
\usepackage[caption=false]{subfig}
\usepackage[strict]{changepage}
\usepackage{xcolor}
\usepackage{ulem}
\usepackage[colorlinks=true,citecolor=blue]{hyperref}
\usepackage{float}
\usepackage{amssymb}
\usepackage{amsmath}
\usepackage{lscape}
\usepackage{mathtools}
\usepackage{natbib}
\usepackage{pdflscape}
\usepackage{gensymb}

\defcitealias{zhou2023}{Paper~I}   

\DeclareSIUnit\jansky{Jy}

\newcolumntype{L}[1]{>{\raggedright\let\newline\\\arraybackslash\hspace{0pt}}m{#1}}
\newcolumntype{C}[1]{>{\centering\let\newline\\\arraybackslash\hspace{0pt}}m{#1}}
\newcolumntype{R}[1]{>{\raggedleft\let\newline\\\arraybackslash\hspace{0pt}}m{#1}}

\begin{document}

\title{Feedback from protoclusters does not significantly change the kinematic properties of the embedded dense gas structures}
\author{J. W. Zhou\inst{\ref{inst1}} 
\and S. Dib\inst{\ref{inst2}}
\and F. Wyrowski\inst{\ref{inst1}}
\and T. Liu\inst{\ref{inst3}}
\and S. H. Li\inst{\ref{inst2}}
\and P. Sanhueza\inst{\ref{inst9},\ref{inst14}}
\and M. Juvela\inst{\ref{inst4}}
\and F. W. Xu\inst{\ref{inst5},\ref{inst6}}
\and H. L. Liu\inst{\ref{inst7}}
\and T. Baug\inst{\ref{inst11}}
\and Y. P. Peng\inst{\ref{inst8}}
\and K. M. Menten\inst{\ref{inst1}}
\and L. Bronfman\inst{\ref{inst10}}
\and C. W. Lee\inst{\ref{inst12},\ref{inst13}}
}
\institute{Max-Planck-Institut f\"{u}r Radioastronomie, Auf dem H\"{u}gel 69, 53121 Bonn, Germany \label{inst1} \\
\email{jwzhou@mpifr-bonn.mpg.de}
\and
Max Planck Institute for Astronomy, K\"{o}nigstuhl 17, 69117 Heidelberg, Germany \label{inst2}
\and
Shanghai Astronomical Observatory, Chinese Academy of Sciences, 80 Nandan Road, Shanghai 200030, Peoples Republic of China \label{inst3}
\and
Department of Physics, P.O.Box 64, FI-00014, University of Helsinki, Finland \label{inst4}
\and
Kavli Institute for Astronomy and Astrophysics, Peking University, Haidian District, Beijing 100871, People’s Republic of China \label{inst5}
\and
Department of Astronomy, School of Physics, Peking University, Beijing 100871, People’s Republic of China \label{inst6}
\and
Department of Astronomy, Yunnan University, Kunming, 650091, PR China \label{inst7}
\and
Department of Physics, Faculty of Science, Kunming University of Science and Technology, Kunming 650500, People's Republic of China \label{inst8}
\and
National Astronomical Observatory of Japan, National Institutes of Natural Sciences, 2-21-1 Osawa, Mitaka, Tokyo 181-8588, Japan \label{inst9}
\and
Departamento de Astronomía, Universidad de Chile, Casilla 36-D, Santiago, Chile
\label{inst10}
\and
Satyendra Nath Bose National Centre for Basic Sciences, BlockJD, Sector-III, Salt Lake, Kolkata-700 106
\label{inst11}
\and
Korea Astronomy and Space Science Institute, 776 Daedeokdaero, Yuseonggu, Daejeon 34055, Republic of Korea
\label{inst12}
\and
University of Science and Technology, Korea (UST), 217 Gajeong-ro, Yuseong-gu, Daejeon 34113, Republic of Korea
\label{inst13}
\and
Astronomical Science Program, The Graduate University for Advanced Studies, SOKENDAI, 2-21-1 Osawa, Mitaka, Tokyo 181-8588, Japan
\label{inst14}
}

\date{Accepted XXX. Received YYY; in original form ZZZ}

\abstract
{A total of 64 ATOMS sources at different evolutionary stages were selected to investigate the kinematics and dynamics of gas structures under feedback. We identified dense gas structures based on the integrated intensity map of H$^{13}$CO$^+$ J=1-0 emission, and then extracted the average spectra of all structures to investigate their velocity components and gas kinematics. For the scaling relations between velocity dispersion $\sigma$, effective radius $R$ and column density $N$ of all structures,
$\sigma-N*R$ always has a stronger correlation compared to $\sigma-N$ and $\sigma-R$. 
There are significant correlations between velocity dispersion and column density, which may imply that the velocity dispersion originates from gravitational collapse, also revealed by the velocity gradients.
The measured velocity gradients for dense gas structures in early-stage sources and late-stage sources are comparable, indicating gravitational collapse through all evolutionary stages. Late-stage sources do not have large-scale hub-filament structures,  but the embedded dense gas structures in late-stage sources show similar kinematic modes to those in early- and middle-stage sources. These results may be explained by the multi-scale hub-filament structures in the clouds.
We quantitatively estimated the velocity dispersion generated by the outflows, inflows, ionized gas pressure and radiation pressure, and found that the ionized gas feedback is stronger than other feedback mechanisms. However,
although feedback from H{\sc ii} regions is the strongest, it does not significantly affect the physical properties of the embedded dense gas structures.
Combining with the conclusions in \citet{Zhou2023arXiv} on cloud-clump scales, we suggest that although
feedback from cloud to core scales will break up the original cloud complex, the substructures of the original complex can be reorganized into new gravitationally governed configurations around new gravitational centers. This process is accompanied by structural destruction and generation, and changes in gravitational centers, but gravitational collapse is always ongoing.}


\titlerunning{Kinematic properties of dense gas structures under feedback}
\authorrunning{J. W. Zhou}

\maketitle 

\section{Introduction}\label{intro}
Stellar feedback from high-mass stars (M$\textgreater$8 M$_{\odot}$) can strongly influence their surrounding interstellar medium and regulate star formation.
The combined effects of multiple feedback mechanisms may play an major role in determining star formation efficiency and the initial mass function in molecular clouds \citep{Dib2010-405,Dib11-415, Krumholz2014-243K}. In \citet{Krumholz2014-243K}, feedback mechanisms are divided into three categories: (1) Momentum feedback, including protostellar outflows and (probably) radiation pressure; (2) Explosive feedback, including winds from hot main sequence stars, photoionizing radiation, and supernovae; (3) Thermal feedback, with non-ionizing radiation as the main form. \citet{Lopez2014-795} assessed the role of stellar feedback at the scales $\sim$10–100 pc toward 32 H{\sc ii} regions in the Large and Small Magellanic Clouds, they employed multi-wavelength data to examine several stellar feedback mechanisms, and found that the ionized gas feedback dominates over the other mechanisms in all of the sources. Similar methods were also used in \citet{Olivier2021-908}, but for 106 deeply embedded H{\sc ii} regions with radii < 0.5 pc. They concluded that the dust-processed radiation dominates in 84\% of the samples.
In simulations, massive stellar feedback, including ionizing radiation, stellar winds, and supernovae \citep{Matzner2002-566, Dale2012-424, Rogers2013-431, Dale2014-442, Rahner2017-470, Smith2018-478, Lewis2023-944}, can suppress the star formation and destroy the natal cloud. However, whether stellar feedback promotes or suppresses star formation remains controversial. 
The simulations of \citet{Dale2014-442} examined the effects of photoionization and momentum-driven winds from O-stars on molecular clouds. They found that feedback has little effect on more massive and denser clouds and it mainly destroys clouds with lower mass and density. \citet{Dib13-436} showed, using semi-analytical models in which gas-expulsion from protocluster clouds is driven by stellar wind feedback originating from both B and O stars that the gas expulsion timescale depends on the clouds mass and the coupling efficiency of the wind with the surrounding gas.   
The observations of \citet{
Pabst2019-565} and \citet{Henshaw2022-509} suggest
that at early stages, energy-winds could dominate, even if this is questioned by
recent theoretical works \citep{
Lancaster2021-914,Lancaster2021-922}.
It is also possible for the feedback to be positive by triggering new star formation. 
In \citet{Deharveng2005-433}, \citet{Zavagno2006-446} and \citet{Liu2017-602}, using data at various wavelengths, they identified and studied
several objects in which star formation appears to have been triggered at the borders of H{\sc ii} regions. \citet{Thompson2012-421} studied a large sample of infrared bubbles, and estimated that the fraction of massive stars in the Milky Way formed by triggering could be between 14 and 30\%. \citet{Kendrew2012-755} found that approximately $\sim$22\% of massive young stars may have formed as a result of feedback from expanding H{\sc ii} regions.

As presented in \citet{Watkins2019-628A}, stellar feedback from O stars clearly modify the structure of the larger scale clouds, but without much effect on the dynamical properties of the assembled dense gas.
In the G305 GMC, feedback has triggered star formation by the collect and collapse mechanism based on observations using the Large APEX sub-Millimeter Array (LAsMA) 7 beam receiver on the Atacama Pathfinder Experiment 12 meter submillimeter telescope (APEX)
 \citep{Mazumdar2021-650,Mazumdar2021-656}. 
\citet{Rugel2019-622} showed that feedback from the first formed cluster in W49A is not strong enough to disperse the cloud, which likely only affect limited parts of W49A. Moreover, all feedback models used in \citet{Rugel2019-622} predict recollapse of the shell after the first star formation event.

The importance of large-scale gravitational collapse in star formation regions has been suggested in many observational and theoretical work \citep{
Hartmann2007-654,Vazquez2009-707,Schneider2010-520,Peretto2013-555,Peretto2014-561,Wyrowski2016-585,Hacar2017-602,Vazquez2019-490,Xu2023ATOMS-XV}. 
In a sample of 11 massive clumps, high-resolution data reveal that protoclusters evolves to tighter by global gravitational collapse \citep{Xu2023ASSEMBLE}.
In the LAsMA observation for the G333 complex under feedback, on large scales, \citet{Zhou2023-676,Zhou2023arXiv} found that the inflow may be driven by the larger-scale structure, consistent with the hierarchical structure in the molecular cloud and gas inflow from large to small scales. The large-scale gas inflow is driven by gravity, implying that the molecular clouds in the G333 complex may be in a state of global gravitational collapse, also refer to \citet{Dib2023-524}. Ubiquitous density and velocity fluctuations also imply the widespread presence of local gravitational collapse \citep{Henshaw2020-4}. 
For local dense gas structures, most of them have obvious correlation between velocity dispersion and density, which indicates the gravitational origin of velocity dispersion. For the G333 complex, although the original cloud complex is disrupted by feedback, the substructures of the original complex can be reorganized into new gravitationally governed configurations around new gravitational centers. This process is accompanied by structural destruction and generation and changes in gravitational centers, but gravitational collapse is always ongoing.
The results in \citet{Zhou2022-514,Zhou2023-676,Zhou2023arXiv} and \citet{Peretto2023-525} show that the kinematic properties of parsec-scale clumps in two very different physical environments (infrared dark and infrared bright) are comparable. Thus, feedback in infrared bright star-forming regions, such as the G333 complex, will not change the kinematic properties of parsec-scale clumps, 
which is also consistent with the survey results that most Galactic parsec-scale massive clumps seem to be gravitationally bound regardless of their evolutionary phases \citep{Liu2016-829, Urquhart2018-473, Evans2021-920, Dib2023-524}. 

In this work, we extend the cloud-clump analysis in \citet{Zhou2023arXiv} to the clump-core scale using high-resolution ALMA data from the ATOMS (ALMA Three-millimeter Observations of Massive Star-forming regions) survey \citep{Liu2020}. 
The ATOMS survey 
has observed 146 active star forming regions at ALMA Band 3. The data can be used to systematically investigate the spatial distribution of various dense gas tracers in a large sample of Galactic massive clumps.
The dense molecular gas structures inside the cloud serve as star-forming sites, their physical states  directly determine the star formation capability of the molecular cloud under feedback. A more systematic study is needed to evaluate how massive protostars influence the dense gas distribution and star formation efficiency in their natal clumps. Using ALMA data, we zoom in the dense gas structures close to or even connecting with protoclusters, the effects of feedback are then inferred through the physical properties of the surrounding dense gas structures.
The sample of the ATOMS survey includes 146 active star forming regions. In \citet{Zhou2022-514}, we found that hub-filament systems are very common within massive protoclusters, and stellar feedback from HII regions gradually destroys the hub-filament systems as protoclusters evolve. The sources at different evolutionary stages are dominated by different feedback mechanisms. By comparing the physical properties of dense gas structures in the clumps at different evolutionary stages, we will be able to infer the relative importance of various feedback mechanisms.  

\section{Observations} 
\label{sec:obs}

\subsection{ALMA observations}
\label{alma}

We use ALMA data from the ATOMS survey (Project ID: 2019.1.00685.S; PI: Tie Liu). 
The ATOMS sources were initially selected from a complete and homogeneous CS J=2-1 molecular line survey toward IRAS sources with far-infrared colors characteristics of UC H{\sc ii} regions \citep{Bronfman1996-115}. As discussed in \citet{Liu2020}, the ATOMS sources are distributed in very different environments of the Milky Way, and are an unbiased sample of the  proto-clusters with the strongest CS J=2-1 line emission (T$_A>2$ K) located in the inner Galatic plane of $-80$\degr$<l<40$\degr, $|b|<2$\degr \citep{Faundez2004-426}.

For the data reduction,
the details of the 12 m array and 7 m array ALMA observations were summarised in \cite{Liu2020,Liuh2021}. Calibration and imaging were carried out using the CASA software package version 5.6 \citep{McMullin2007}. The 7 m data and 12 m array data were calibrated separately. Then the visibility data from the 7 m and 12 m array configurations were combined and later imaged in CASA. For each source and each spectral window (spw), a line-free frequency range is automatically determined using the ALMA pipeline. This frequency range is used to (a) subtract continuum from line emission in the visibility domain, and (b) make continuum images. Continuum images are made from multi-frequency synthesis of data in the line-free frequency ranges in the two 1.875 GHz wide spectral windows, spws 7 and 8, centered on $\sim99.4$ GHz (or 3 mm). Visibility data from the 12 m and 7 m arrays are jointly cleaned using task tclean in CASA 5.6. We used natural weighting and a multiscale deconvolver, for an optimized sensitivity and image quality. All images are primary-beam corrected. The continuum image reaches a typical 1 $\sigma$ rms noise of $\sim$0.2 mJy with a synthesized beam FWHM size of $\sim2.2\arcsec$. 
In this work, we also use H$^{13}$CO$^+$ J=1-0 (86.754288 GHz) and HCO$^+$ J=1-0 (89.188526 GHz) molecular lines data with spectral resolutions of 0.211 km~s$^{-1}$ and 0.103 km~s$^{-1}$. The typical beam FWHM size and channel rms noise level for H$^{13}$CO$^+$ J=1-0 line emission are $\sim2.5\arcsec$ and 8 mJy~beam$^{-1}$, respectively. For HCO$^+$ J=1-0 line emission,
the two values are $\sim2.4\arcsec$ and 12 mJy~beam$^{-1}$.
The typical maximum recovered angular scale in this survey is about 1 arcmin, which is comparable to the field of view (FOV) in the 12 m array observations \citep{Liu2020}. 

\subsection{Spitzer infrared data }
We also use images at 8 $\mu$m, obtained by the Spitzer Infrared Array Camera (IRAC), as part of the GLIMPSE project \citep{Benjamin2003}. The images of IRAC were retrieved from the Spitzer Archive and the angular resolution of the images are about $2\arcsec$.

\section{Results}

\subsection{Evolutionary stages}\label{classify}

\begin{figure}
\centering
\includegraphics[width=0.45\textwidth]{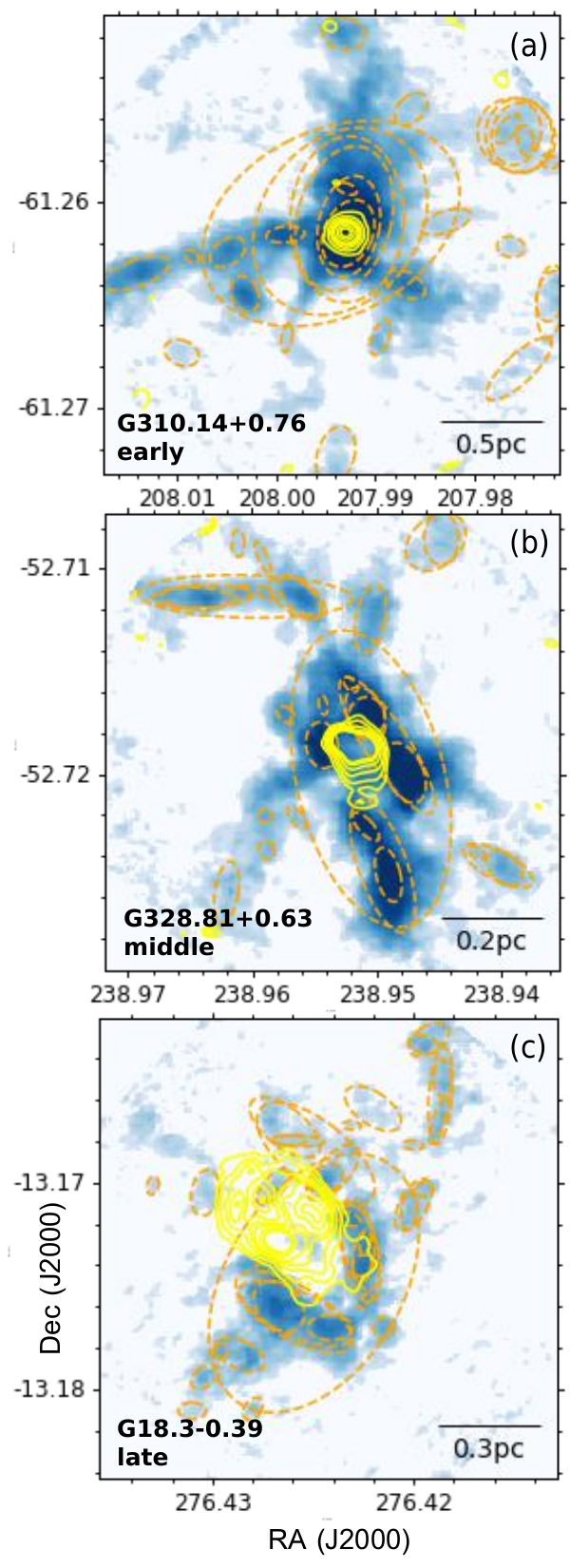}
\caption{The Moment 0 maps of H$^{13}$CO$^{+}$ J=1-0 for 3 ATOMS sources at three evolutionary stages. Yellow contours show 3 mm continuum emission with its minimum contour level of 5$\sigma$. Orange eclipses represent the retained Dendrogram structures.
The maps for all 64 ATOMS sources studied in this work are presented in Appendix.\ref{all}.}
\label{map}
\end{figure}

\begin{figure}
\centering
\includegraphics[width=0.45\textwidth]{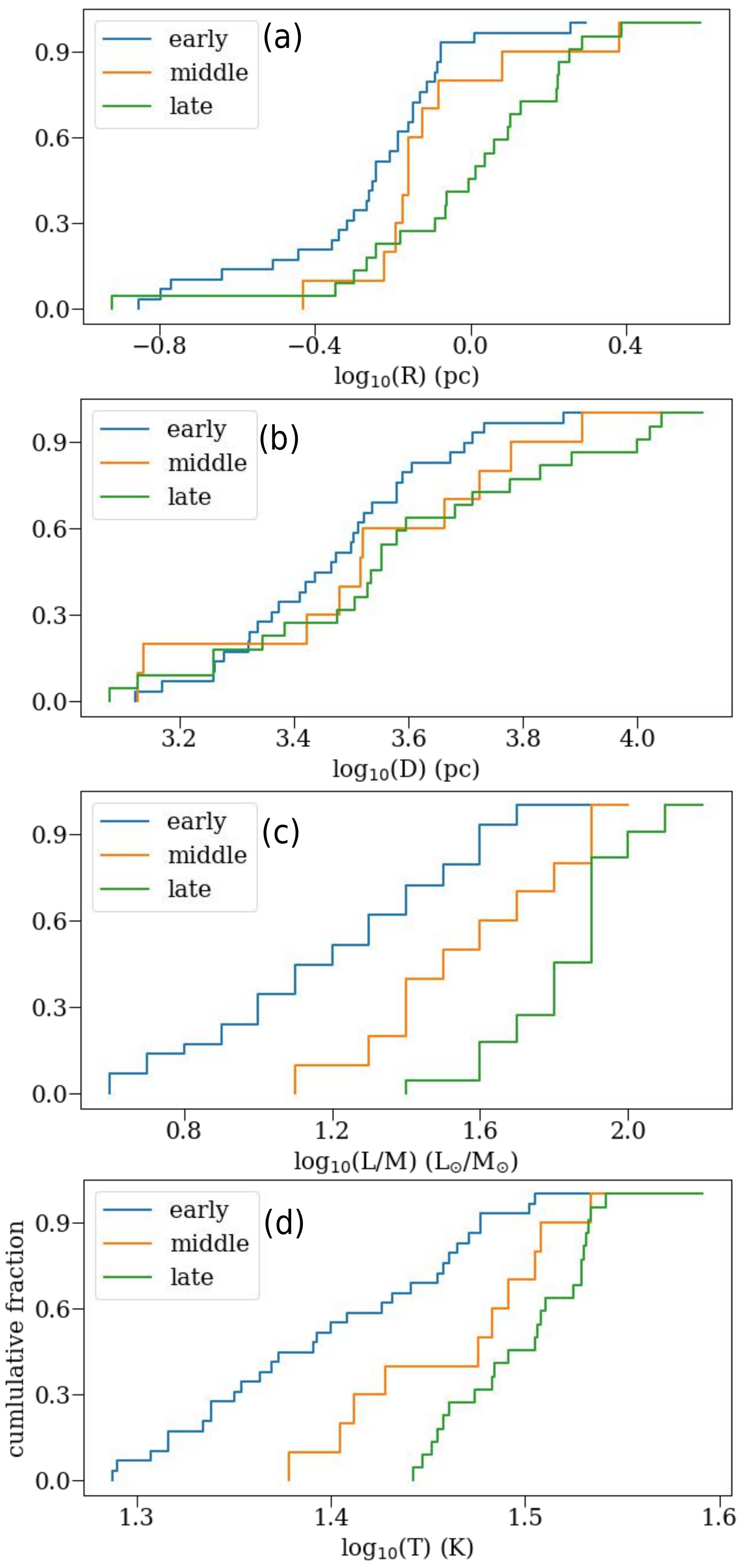}
\caption{Physical parameters of 64 ATOMS sources at different evolutionary stages. Panels (a), (b), (c) and (d) show the distributions of the radius $R$, distance $D$, luminosity-to-mass ratios $L/M$ and dust temperature $T$, respectively.}
\label{evolution}
\end{figure}

Generally, the infrared properties and the presence of H40$_{\rm \alpha}$ emission can be used to assess the evolutionary stages of the sources \citep{Chambers2009-181,Sanhueza2012-756}.
As discussed in \citet{Zhou2022-514}, stellar feedback from H{\sc ii} regions will gradually destroy the hub-filament systems as protoclusters evolve. Thus, the morphology of the hub-filament structure can also characterize the evolutionary stages.
In Fig.2 and Fig.3 of \citet{Zhou2022-514}, for hub-filament systems, the overlaid contours of 3 mm continuum should be located in the center of filamentary network with the strongest H$^{13}$CO$^+$ J=1-0 emission, where the center of the gravitational potential well is located and is the potential site for high-mass star formation. In the late stage of evolution, the 3 mm continuum emission is almost completely separated from H$^{13}$CO$^+$ J=1-0 emission, indicating that the relatively evolved H{\sc ii} regions have dispersed their natal gas, also refer to \citet{Liu2023-522}. Thus the separation between 3mm continuum emission and H$^{13}$CO$^{+}$ (1-0) emission can also be used for the classification. Finally, we divided ATOMS sources into three evolutionary stages using the following criteria: 

1. The sources in the early stage (early-sources): (1) Infrared dark and without H40$_{\rm \alpha}$ emission; (2) Well-defined hub-filament  structures; (3) Their 3mm continuum emissions concentrate on the centers of filamentary networks with the strongest H$^{13}$CO$^{+}$ (1-0) emission.

2. The sources in the middle stage (middle-sources) have the same features like early-sources except for having compact H40$_{\rm \alpha}$ emission. 

3. The sources in the late stage (late-sources): (1) Obvious triggering signs, such as the bent dense gas structures traced by H$^{13}$CO$^{+}$ (1-0) emission; (2) Their 3mm emission is extended, and all of them have strong H40$_{\rm \alpha}$ and 8 $\mu$m emission, indicating that 3mm continuum emission is dominant by free–free emission \citep{Keto2008-672, Zhang2023-520}. 
The two conditions mean that hubs in the late stage have developed into protoclusters. Then powerful H{\sc ii} regions driven by protoclusters will destroy the hub-filament systems eventually. We can see the expansion of H{\sc ii} regions are tearing up the surrounding gas structures in Fig.~\ref{map}(c). 

The number of early-sources, middle-sources and late-sources are 30, 11, and 23. We have applied stringent criteria for the selection, and retained only sources that can be clearly catergorized in one out of the three stages. The values of radius, distance, bolometric luminosity, mass, and dust temperature for ATOMS sources have been listed in the Table.A1 of \citet{Liu2020}. The luminosity-to-mass ratios $L/M$ and dust temperature $T$ of the clumps can be used to trace the evolutionary process of star formation \citep{Saraceno1996-309,Krumholz2007-654,Molinari2008-481,Liu2016-829, Stephens2016-824,Urquhart2018-473}. 
As shown in Fig.\ref{evolution}(c) and (d), from early to middle to late stages, $L/M$ and $T$ of the sources are indeed increasing gradually, confirming that our classification is reasonable. 

\subsection{Dendrogram structures}\label{Dendrogram}

\begin{figure}
\centering
\includegraphics[width=0.5\textwidth]{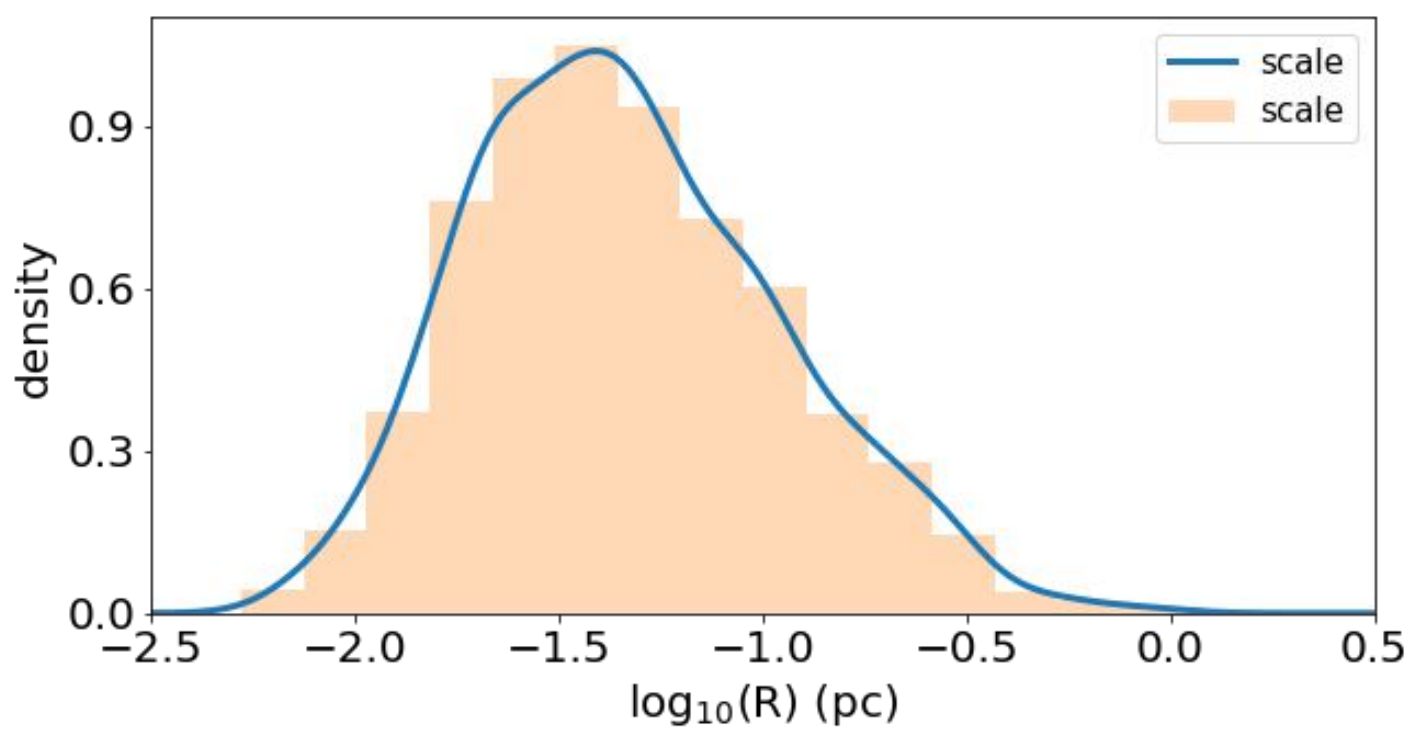}
\caption{Scale distribution of all identified structures. The probability density is estimated by the kernel density estimation (KDE) method.}
\label{scale}
\end{figure}

The issues of identifying the structures in PPV cube by the Dendrogram algorithm have been described in Sec.3.1 and Sec.3.2 of \citet{Zhou2023arXiv}. Thus, we adopt the same method in \citet{Zhou2023arXiv} to identify dense gas structures. We directly identify hierarchical (sub-)structures based on the 2D integrated intensity (Moment 0) map of H$^{13}$CO$^{+}$ (1$-$0) emission, and then extract the average spectrum of each structure to investigate their velocity components and gas kinematics. 

For the Moment 0 map, a 5$\sigma$ threshold has been set, we therefore only require the smallest area of the structure to be larger than 0.5 beam and do not set other parameters to reduce the dependence of the identification on the algorithm parameters. In Fig.\ref{map}, the structures identified by Dendrogram correspond well to the background integrated intensity maps.


The algorithm will approximate the morphology of each structure as an ellipse. In Dendrogram, the long and short axes of an ellipse, $a$ and $b$, are the rms sizes (second moments) of the intensity distribution along the two spatial dimensions. 
As described in \citet{Zhou2023arXiv}, $a$ and $b$ will give a smaller ellipse, compared to the size of the identified structure, thus multiplying a factor of 2 to enlarge the ellipse is necessary. Then the effective physical radius of an ellipse is $ R\rm_{eff}$ =$\sqrt{2a \times 2b}*D$, where $D$ is the distance of the source.

As discussed in \citet{Zhou2023arXiv}, two screening criteria were used to filter the structures:
(1) Eliminating the repetitive branch structures;
(2) Excluding branch structures with complex morphology. Then
for 2477 retained structures, according to their averaged spectra, 57 structures with absorption features are eliminated.
Then following the procedure in \citet{Zhou2023arXiv}, 
we fitted the averaged spectra of 2420 structures individually using the fully automated Gaussian decomposer \texttt{GAUSSPY+} \citep{Lindner2015-149, Riener2019-628} algorithm. The parameter settings of the decomposition are the same as that in \citet{Zhou2023-676}. According to the line profiles, all averaged spectra are divided into three categories: type1 (single velocity component, 2038), type2 (separated velocity components, 42) and type3 (blended velocity components, 340), also refer to the Fig.6 of \citet{Zhou2023arXiv} for the examples of the typical line profiles.
In the subsequent analysis, we discard type2 structures because they are few (1.7\%) to simplify the discussion. Furthermore these structures are likely the product of overlapping of unrelated gas components.

In this work, we do not classify the structures by the nomenclatures used in {\it astrodendro}\footnote{\url{https://dendrograms.readthedocs.io/en/stable/index.html}}, such as leaf, branch or trunk, because there is always a considerable overlap of the scales between leaf and branch structures, as described in \citet{Zhou2023arXiv}. Moreover, in Fig.\ref{scale}, the scales of all structures are presented as a continuous unimodal distribution.


\subsection{Column density}\label{column}
To derive the column densities from H$^{13}$CO$^{+}$ (1-0) emission, we assume conditions of local thermodynamic equilibrium (LTE) and a beam filling factor of 1. The LTE analysis is described in detail in Appendix.\ref{lte-e}. 
In \citet{Shimajiri2017-604}, for the Aquila, Ophiuchus, and Orion B clouds,
the H$^{13}$CO$^+$ abundances relative to H$_2$,
$X_{\rm H^{13}CO^+} \equiv N_{\rm H^{13}CO^+}/N_{\rm H_2}$, have mean values in the range (1.5-5.8)$\times$10$^{-11}$. 
These abundance estimates are consistent within a factor of a few 
with the findings in other regions, such as 1.1$\pm$0.1$\times$10$^{-11}$ in OMC 2-FIR 4 \citep{Shimajiri2015-217} and 1.8$\pm$0.4$\times$10$^{-11}$ in Sagittarius A \citep{Tsuboi2011-63}. 
\citet{Peretto2013-555} derived a H$^{13}$CO$^+$ abundance of $5\times10^{-11}$ from Mopra observations towards SDC335 using the 1D non-LTE RADEX radiative transfer code \citep{van-Tak2007-468}. 
Here we take the mean value $X_{\rm H^{13}CO^+} \sim 4.2\times10^{-11}$ shown in Table 2 of \citet{Shimajiri2017-604}.

\subsection{Velocity components}

In \citet{Zhou2023arXiv}, type3 structures were discarded because of the complex line profiles. In this work, in order to increase the sample of each type of sources, we need to consider type3 structures. The possible reasons for the complex profiles are : (1) the optical depth effect; (2) the overlapping of uncorrelated gas components; (3) the complex gas kinematics inside the structures. Generally, H$^{13}$CO$^+$ J=1-0 line emission is optically thin, thus we do not consider the first factor. The second factor has been discussed in detail in \citet{Zhou2022-514},
it is not an issue in the ATOMS survey, thus our analysis in previous work can be done based on the Moment maps. 
The significant overlap for the line profile of a type3 structure may indicate the correlation between different velocity components in the structure. The more severe the overlap between different velocity components, the more likely they belong to the same structure. In the extreme case, they merge into a single-peak line profile. If the velocity differences between different velocity components are small, these velocity components are likely to be different parts of the same structure rather than unrelated overlapping structures, thus the velocity decomposition is not necessary. For example, in a hub-filament structure, the velocity of gas flow in each filament is different, leading to different velocity components in the same structure. 
Considering the multi-scale hub-filament scenario described in \citet{Zhou2022-514},
type3 structures may be hub-filament structures at different scales. 

In the subsequent analysis,
we treat type3 structures as independent structures like type1 structures, but with more complex gas motions. We will constantly provide evidence to justify this assumption. 
If different velocity components in a type3 structure are different parts of the same structure, the velocity dispersion of the type3 structure can be obtained by an intensity-weighted average for the velocity dispersion of all velocity components within this structure.

\subsection{Statistics of the basic physical quantities}\label{basic}
\begin{figure*}
\centering
\includegraphics[width=1\textwidth]{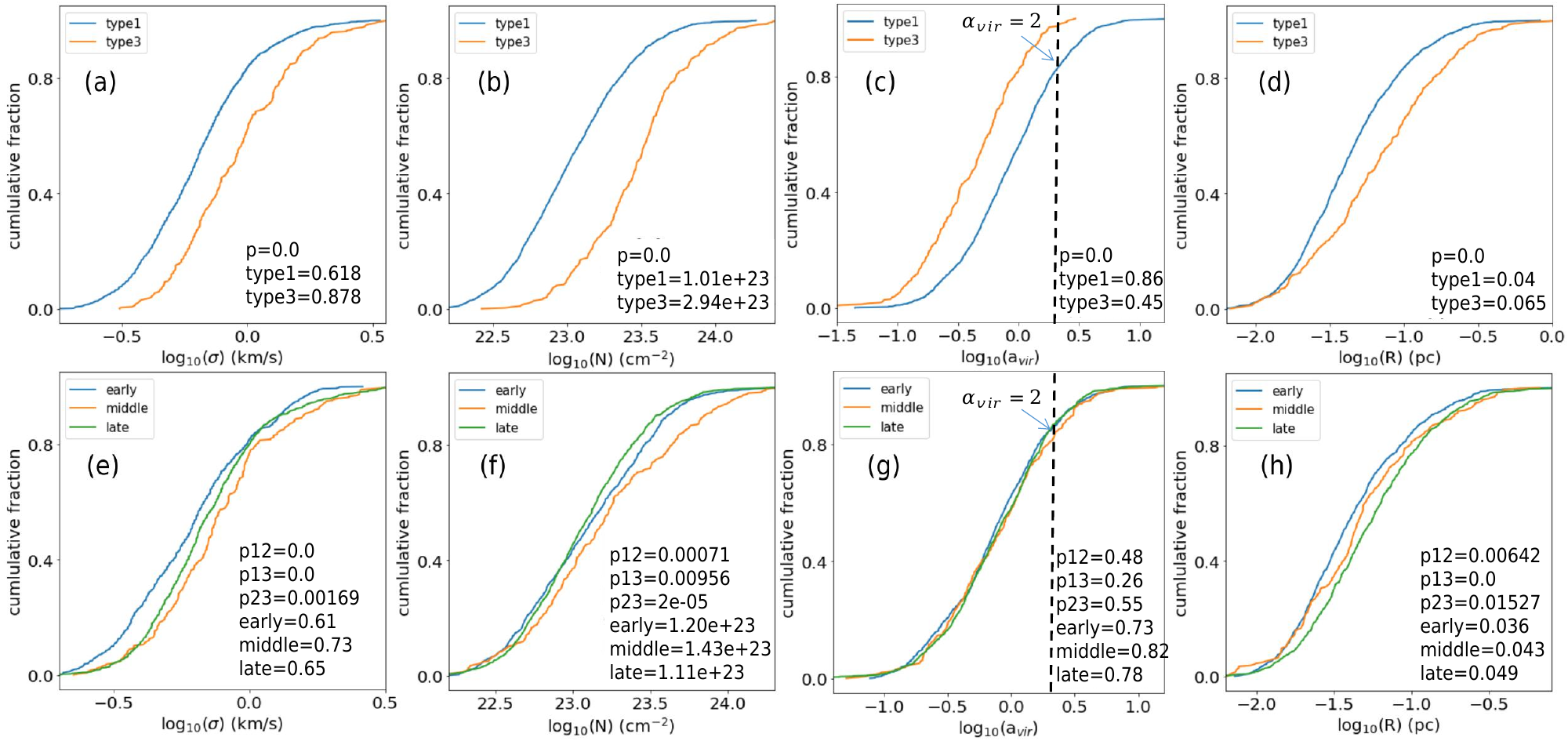}
\caption{Physical properties of different structure types at different evolutionary stages. 
The first row is the comparison between type1 and type3 structures. The second row is the comparison between early-, middle- and late-sources.
Dashed lines in panels (c) and (g) mark the positions of $\alpha_{vir}$=2.
"p" is the p-value of Kolmogorov-Smirnov (KS) test.
"p12" , "p13" and "p23" are the p-values of KS tests between the structures inside early- and middle-sources, early- and late-sources, middle- and late-sources, respectively. 
The median values also show in the legends.}
\label{quantity}
\end{figure*}

Below, we discuss the differences in the physical properties of different structure types at different evolutionary stages.
For two data sets, the magnitude comparison is based on their median values. 
Generally, the p-value of their Kolmogorov-Smirnov (KS) test is far less than 0.05, if their median values are significantly different, as shown in Fig.\ref{quantity}. 

From the first line of Fig.\ref{quantity},
type3 structures have larger velocity dispersion, higher column density and lower virial ratios than type1 structures at different evolutionary stages, 
indicating that
type3 structures mainly concentrate on the hub regions, thus have higher column density and present multiple velocity components. In \citet{Zhou2022-514}, from the grid maps of H$^{13}$CO (1$-$0) emission, the hub regions
always show multiple velocity components which should be attributed to the complex gas motions, not the superposition of uncorrelated foreground or background velocity components. 

From the second line of Fig.\ref{quantity},
although late-sources have powerful H{\sc ii} regions, their inner gas structures do not have significantly larger velocity dispersion and virial ratios, compared to that in early- and middle-sources. In Fig.\ref{quantity}(e), the embedded dense structures inside middle-sources even have slightly larger velocity dispersion than those inside late-sources.
These statistical results are unusual 
and we need to reconsider the effects of feedback from expanding H{\sc ii} regions, see Sec.\ref{feedback} for more discussion. Moreover, the higher column density of the embedded dense structures inside middle-sources in Fig.\ref{quantity}(f) keeps in line with the gravitational collapse of hub-filament systems.

\subsection{Scaling relations}\label{column}

\begin{figure*}
\centering
\includegraphics[width=0.95\textwidth]{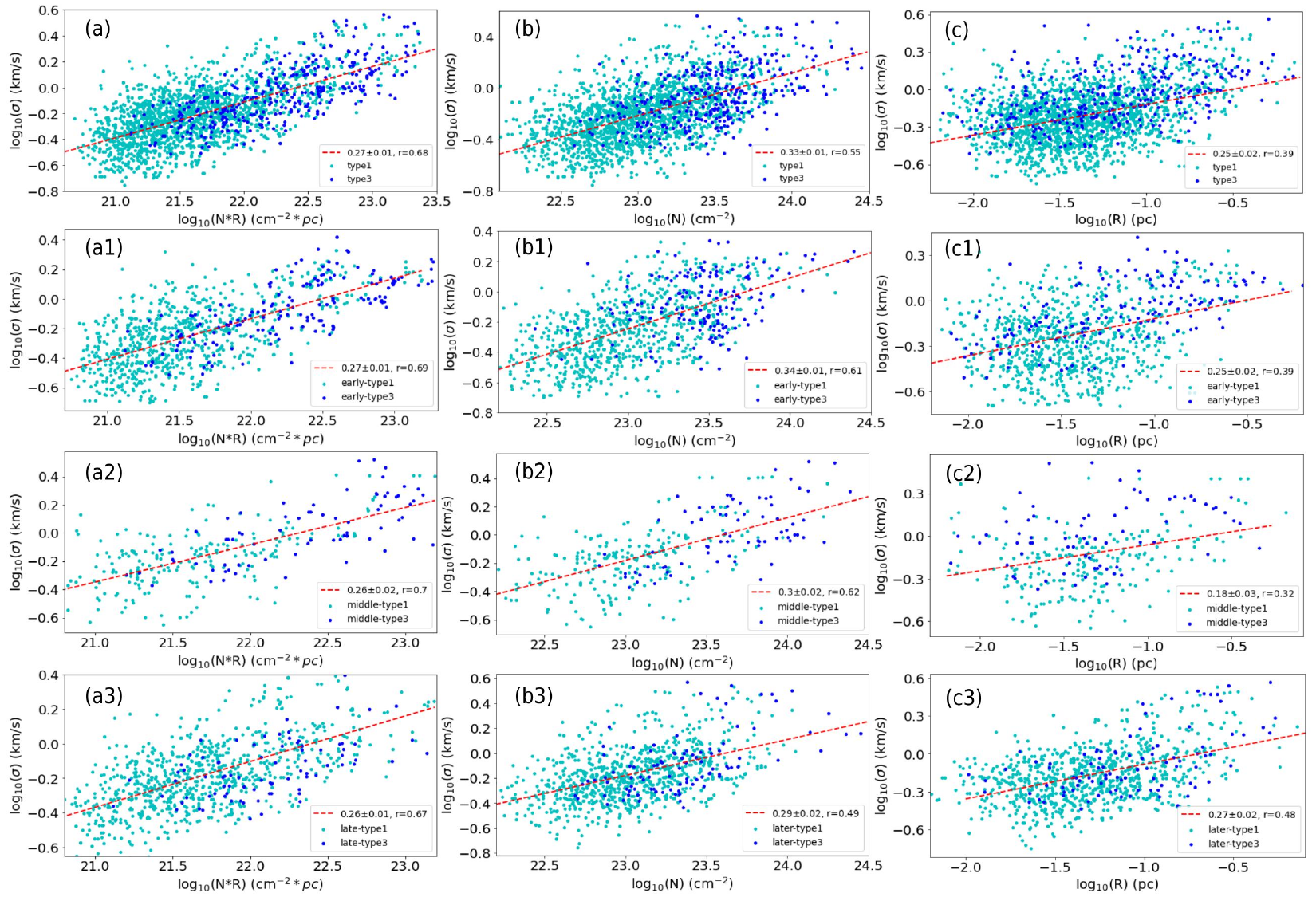}
\caption{Scaling relations of type1 and type3 structures. (a) $\sigma-N*R$ (b) $\sigma-N$; (c) $\sigma-R$. $\sigma$, $R$ and $N$ are the velocity dispersion, effective radius and column density of each structure, respectively. 'r' represents the Pearson correlation coefficient. The values in the legends are the exponents of the power laws.
The second, third and fourth rows are the same as the first row, but for early-, middle- and late-sources, respectively.}
\label{scaling}
\end{figure*}

The scaling relations between different quantities contain information on the physical conditions of the structures. The scaling relations between velocity dispersion $\sigma$, effective radius $R$ and column density $N$ shown in Fig.\ref{scaling} are similar to those in \citet{Zhou2023arXiv}.
$\sigma-N*R$ always has stronger correlation compared to $\sigma-N$ and $\sigma-R$. 
\footnote{For a more convenient comparison with $\sigma-R$ and $\sigma-N$ relations, we convert the Heyer relation $\sigma/R^{0.5} \propto N^{0.5}$ to the form $\sigma \propto (R*N)^{0.5}$ \citep[Eq.\,3 in][]{Ballesteros2011-411}, both should have a slope of 0.5.}
However, the slopes of $\sigma-N*R$ relations are $\sim$0.27, which may indicate the slowing down of pure
free-fall gravitational collapse.
There are significant correlations between velocity dispersion and column density, which may imply that the velocity dispersion originates from the gravitational collapse, also revealed by the measured velocity gradients in Sec.\ref{gradient-s}.
$\sigma-R$ has a poor correlation,
that is, a significant deviation from the Larson-relation \citep{Larson1981-194}.

Additionally, it is noteworthy that both type1 and type3 structures present similar scaling relations, which further confirm that type3 structures are independent like type1 structures. The structures generated by random overlap should not show good scaling relations.

\subsection{Velocity gradient}\label{gradient-s}

\begin{figure}
\centering
\includegraphics[width=0.45\textwidth]{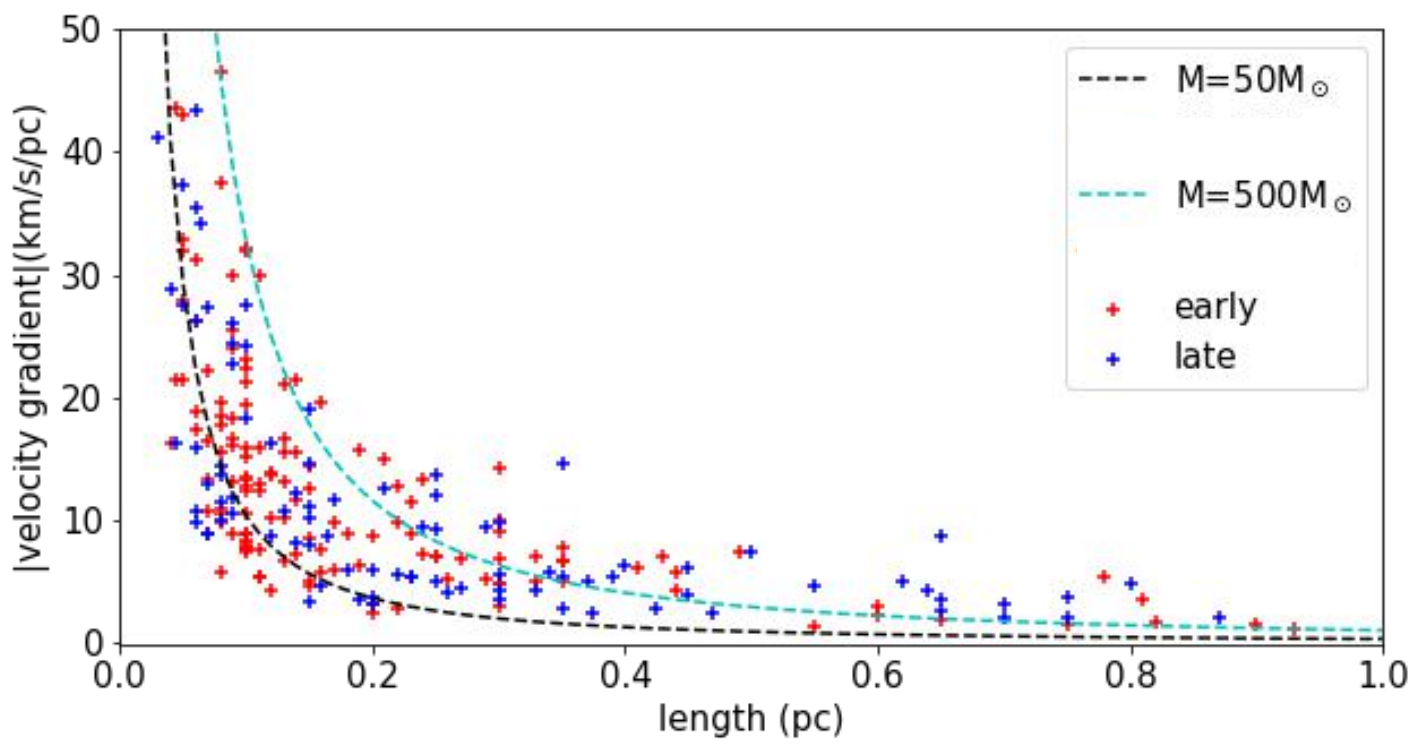}
\caption{Velocity gradient versus the length over which the gradient has been fitted. Blue "+" and red "+" represent the velocity gradients fitted in early-sources and late-sources. The dashed lines show free-fall velocity gradients for comparison. For the free-fall model, black and cyan lines denote masses of 50M$_\odot$ and 500M$_\odot$, respectively.}
\label{gradient}
\end{figure}

In \citet{Zhou2022-514}, we used the Moment 0 maps (integrated intensity maps) of H$^{13}$CO$^{+}$ J=1-0 to identify filaments in ATOMS sources, and extracted the velocity and intensity along the filament skeletons from the Moment 1 and Moment 0 maps.  
Clear velocity and density fluctuations are seen along the filaments, allowing us to fit velocity gradients around the intensity peaks. 
The filaments like the streamlines, which can trace  how do gas inflows converge to local dense structures. Local dense structures as gravitational centers will accrete the surrounding diffuse gas and then form local hub-filament structures. A hub-filament structure can be translated into a funnel structure in PPV space as described in \citet{Zhou2023-676}. As illustrated in Fig.9 of \citet{Zhou2023-676}, the gradient of the funnel profile can reflect the strength of the gravitational field. In \citet{Zhou2022-514} and \citet{Zhou2023-676}, the filaments will pass through multiple local hub-filament structures presented as velocity and density fluctuations. Therefore, 
the local velocity gradients along the filaments can reveal the strength of local gravitational fields. 
By comparing local velocity gradients, we can assess the differences in gravitational states of the embedded dense gas structures among different feedback environments.

For all ATOMS sources, the identified filaments, and corresponding extracted 
velocity and density along the filaments have been shown in \citet{Zhou2022-514}.
We first estimated two overall velocity gradients between the velocity peaks and valleys at the two sides of the center of the gravitational potential well and ignored local velocity fluctuations. Then, we also derived local velocity gradients around local intensity peaks of the H$^{13}$CO$^+$ emission (see Fig.\,6 in \citealt{Zhou2022-514}). 
In Fig.\ref{quantity}(g), the virial ratios of the embedded dense gas structures in late-sources and early-sources are comparable.
In Fig.\ref{gradient}, the measured velocity gradients for the structures in early-sources and late-sources are also comparable, once again demonstrating that feedback rarely affects the physical properties of the embedded dense gas structures.

\subsection{Feedback of  early-sources vs. late-sources}\label{feedback}

\begin{table}
	\centering
	\caption{Velocity dispersion produced by different physical processes in early-sources. v$_{\rm turb, out}$, v$_{\rm in}$ and v$_{\rm rad}$ correspond to the outflows, the inflows and the radiation pressure, respectively.}
	\label{line}
	\begin{tabular}{cccc} 
		\hline
  early-sources	&	v$_{\rm turb, out}$ (km s$^{-1}$)	&	v$_{\rm in}$ (km s$^{-1}$)	&	v$_{\rm rad}$ (km s$^{-1}$)	\\
\hline
G261.65-2.09	&	0.07 	&	0.49	$\pm$		0.16	&	0.28	\\
G294.51-1.62	&	0.11 	&	0.59	$\pm$		0.19	&	0.33	\\
G305.8-0.24	&	0.10 	&	0.86	$\pm$		0.32	&	0.6	\\
G305.89+0.02	&	0.12 	&	0.92	$\pm$		0.33	&	0.28	\\
G307.76+0.35	&	0.03 	&	0.81	$\pm$	0.28	&	0.17	\\
G310.14+0.76	&	0.15 	&	1.26	$\pm$		0.47	&	0.91	\\
G313.58+0.32	&	0.03 	&	0.64	$\pm$		0.21	&	0.33	\\
G313.77-0.86	&	0.10 	&	0.95	$\pm$		0.34	&	0.34	\\
G318.05+0.09	&	0.13 	&	1.23	$\pm$		0.45	&	0.61	\\
G326.47+0.7	&	0.24 	&	0.85	$\pm$		0.3	&	0.24	\\
G327.12+0.51	&	0.00 	&	1.28	$\pm$		0.47	&	0.78	\\
G329.47+0.5	&	0.26 	&	1.11	$\pm$		0.42	&	0.29	\\
G329.41-0.46	&	0.04 	&	1.19	$\pm$		0.5	&	1.19	\\
G335.59-0.29	&	0.11 	&	1.14	$\pm$		0.42	&	0.46	\\
G339.62-0.12	&	0.09 	&	1.22	$\pm$		0.42	&	0.41	\\
G339.88-1.26	&	0.06 	&	0.92	$\pm$	0.33	&	1.04	\\
G341.13-0.35	&	0.01 	&	1.31	$\pm$		0.47	&	0.33	\\
G343.13-0.06	&	0.32 	&	0.94	$\pm$	0.35	&	0.74	\\
G352.63-1.07	&	0.11 	&	0.86	$\pm$		0.28	&	0.35	\\
G6.8-0.26	&	0.04 	&	1.08	$\pm$	0.39	&	0.36	\\
G12.42+0.51	&	0.06 	&	1.11	$\pm$	0.38	&	0.42	\\
G12.91-0.26	&	0.20 	&	1.01	$\pm$		0.4	&	0.5	\\
G14.33-0.64	&	0.09 	&	0.98	$\pm$		0.34	&	0.31	\\
G16.59-0.05	&	0.04 	&	1.07	$\pm$		0.4	&	0.47	\\
G19.36-0.03	&	0.11 	&	1.14	$\pm$		0.41	&	0.22	\\
G19.88-0.53	&	0.14 	&	1.1	 $\pm$	0.41	&	0.28	\\
G25.65+1.05	&	0.13 	&	1.03	$\pm$		0.36	&	0.37	\\
G28.86+0.07	&	0.04 	&	1.6	 $\pm$	0.65	&	0.99	\\
G31.58+0.08	&	0.04 	&	0.98	$\pm$		0.36	&	0.51	\\
G37.43+1.52	&	0.05 	&	0.42	$\pm$		0.15	&	0.48	\\
& & \\
median	&	0.10 	&	1.02	$\pm$		0.37	&	0.39	\\
mean	&	0.10 	&	1	$\pm$		0.37	&	0.49	\\
      \hline
	\end{tabular}
    \label{early}
\end{table}

\begin{table}
	\centering
	\caption{Velocity dispersion produced by different physical processes in late-sources. v$_{\rm HII}$ and v$_{\rm rad}$ correspond to the ionized gas pressure and the radiation pressure, respectively.}
	\label{line}
	\begin{tabular}{ccc} 
		\hline
late-sources	&	v$_{\rm HII}$ (km s$^{-1}$)	&	v$_{\rm rad}$ (km s$^{-1}$)	\\
 \hline
G268.42-0.85	&	1.8	$\pm$	0.14	&	0.61	\\
G305.2+0.03	&	6	$\pm$	0.5	&	1.27	\\
G312.11+0.31	&	4.85	$\pm$	0.43	&	1.02	\\
G316.14-0.51	&	4.44	$\pm$	0.36	&	1.15	\\
G324.2+0.12	&	8.98	$\pm$	0.71	&	1.68	\\
G326.45+0.91	&	5.75	$\pm$	0.54	&	1.16	\\
G326.72+0.61	&	6.14	$\pm$	0.52	&	0.84	\\
G329.47+0.22	&	6.78	$\pm$	0.44	&	0.91	\\
G332.65-0.61	&	5.88	$\pm$	0.42	&	1.16	\\
G333.31-0.37	&	7.23	$\pm$	0.61	&	1.4	\\
G337.12-0.17	&	13.79	$\pm$ 1.12	&	2.77	\\
G340.78-1.02	&	4.08	$\pm$	0.26	&	1.16	\\
G344.22-0.59	&	4.41	$\pm$	0.35	&	0.63	\\
G350.51+0.96	&	8.9	$\pm$	0.65	&	1.06	\\
G350.1+0.08	&	2.98	$\pm$	0.21	&	1.63	\\
G1.13-0.11	&	1.44	$\pm$	0.15	&	1.17	\\
G8.14+0.22	&	8.32	$\pm$	0.67	&	1.2	\\
G9.62+0.19	&	9.5	$\pm$	0.64	&	1.41	\\
G11.94-0.62	&	6.68	$\pm$	0.52	&	0.83	\\
G13.87+0.28	&	5.09	$\pm$	0.38	&	1.4	\\
G18.3-0.39	&	6.8	$\pm$	0.48	&	0.9	\\
G23.95+0.15	&	10.33	$\pm$	0.71	&	1.55	\\
G32.8+0.19	&	16.13	$\pm$	1.35	&	1.96	\\
& & \\
median	&	6.14	$\pm$	0.5	&	1.16	\\
mean	&	6.8	$\pm$	0.53	&	1.26	\\
      \hline
	\end{tabular}
    \label{late}
\end{table}

In the analysis of the kinematics in early-sources, we mainly focus on the outflows and inflows. 
When investigating the gas  kinematics in late-sources, we primarily consider the ionized gas pressure.
Moreover, the radiation pressure exists in all sources at different evolutionary stages. But late-sources must have stronger radiation pressure than early-sources, because they embed luminous ionizing (young) stellar objects. 
We note that inflow is not a feedback mechanism. However, as a line broadening mechanism, it is comparable to other feedback mechanisms.
Below, we quantitatively estimate the velocity dispersion generated by the four physical processes, and then compare them with the statistical results in Sec.\ref{basic}.  

\subsubsection{Outflow}
We assume that the line emission of H$^{13}$CO$^+$ J=1-0 and high velocity emission of HCO$^+$ J=1-0 are optically thin, and the excitation temperature of H$^{13}$CO$^+$ J=1-0 and HCO$^+$ J=1-0 lines are the same and constant in both low-velocity gas and high-velocity outflow gas. Then according to equations (4) and (5) in \citet{Zhou2021-508}, the mass of the outflows can be estimated by: 
\begin{equation}
\label{Mout}
\frac{M_{\rm out}}{M_{\rm gas}}\approx \left[\frac{\rm H^{13}CO^{+}}{\rm HCO^{+}}\right]
\frac{F_{\rm out, HCO^{+}}}{F_{\rm gas, H^{13}CO^{+}}}
\end{equation}
The expression in bracket is the abundance ratio of the molecules, here we take $\left[ \rm H^{13}CO^{+}/ \rm HCO^{+}\right]\sim 1/60$ \citep{Milam2005-634}.
$F_{\rm out, HCO^{+}}$ and $F_{\rm gas, H^{13}CO^{+}}$
are the corresponding total velocity-integrated flux (the sum of all pixels on Moment 0 maps) of the outflows and the dense gas of the clump traced by HCO$^{+}$ (1-0) and H$^{13}$CO$^{+}$ (1-0) emission.
$M_{\rm gas}$ is calculated using the column density map derived by LTE analysis in Sec.\ref{column}. After obtaining the value of $M_{\rm out}$, 
according to the total velocity-integrated flux of the redshifted and blueshifted lobes of the outflows $F_{\rm out, HCO^{+}, r}$ and $F_{\rm out, HCO^{+}, b}$, the masses of two lobes are 
\begin{equation}
M_{\rm out, r}=\frac{F_{\rm out, HCO^{+}, r}}{F_{\rm out, HCO^{+}}} \times M_{\rm out}, 
\end{equation}
\begin{equation}
M_{\rm out, b}=\frac{F_{\rm out, HCO^{+}, b}}{F_{\rm out, HCO^{+}}} \times M_{\rm out}. 
\end{equation}
Assuming that the momentum of the outflows completely transfers to the surrounding dense gas (i.e. this is an upper limit), then the corresponding turbulent velocity $v_{\rm turb, out}$ inside the dense gas caused by the outflows can be estimated from the equation
\begin{equation}
M_{\rm gas} v_{\rm turb, out} = M_{\rm out, r} v_{\rm r} + M_{\rm out, b} v_{\rm b}.
\end{equation}
According to the Moment 1 maps of the two lobes, their average velocities are $v_{\rm r0}$ and $v_{\rm b0}$, the systematic velocity of the clump is $v_{\rm 0}$, then the velocities of the two lobes are
$v_{\rm r}=v_{\rm r0}-v_{\rm 0}$ and $v_{\rm b}=v_{\rm 0}-v_{\rm b0}$.

\subsubsection{Inflow}
\begin{figure}
\centering
\includegraphics[width=0.45\textwidth]{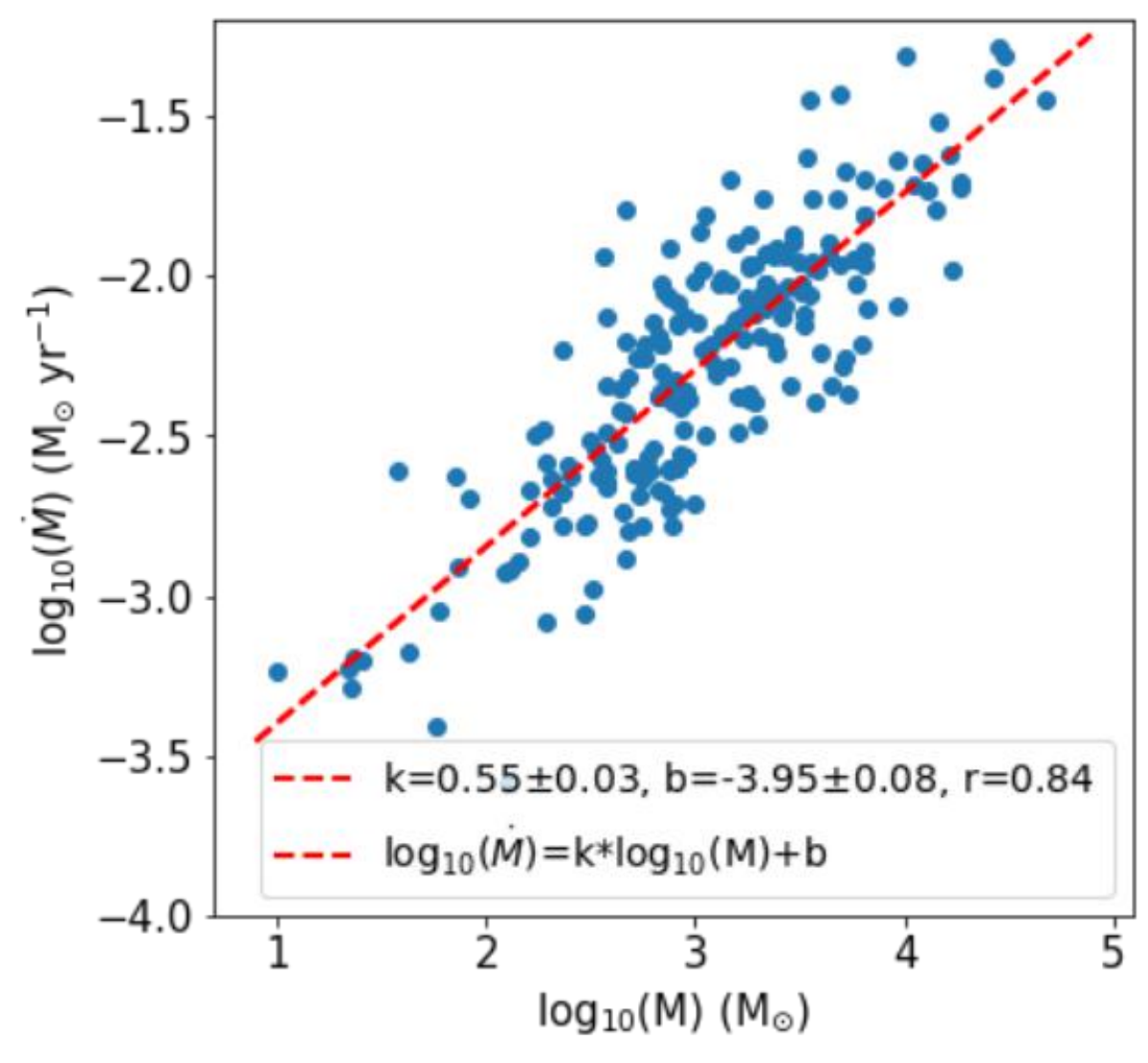}
\caption{The relation between the inflow rate and the mass of the clump. Data points are from the tables of \citet{He2015-450,He2016-461}. "r" is the Pearson coefficient.}
\label{fit}
\end{figure}

Assuming spherical
symmetry, 
the clump has a mass
 inflow rate of
\begin{equation}
\dot M \approx 4 \pi r_{\rm in}^2 \rho v_{\rm in},
\end{equation}
\begin{equation}
\rho =\frac{M_{\rm c}}{4/3\pi R_{\rm c}^{3}}. 
\end{equation}
$R_{\rm c}$ and $M_{\rm c}$ are the radius and mass of the clump, $r_{\rm in}$ and $v_{\rm in}$ are the infall radius and velocity. We note that
the distance $D$, radius $R_{\rm c}$, dust temperature $T_{\rm d}$, mass $M_{\rm c}$ and bolometric luminosity $L_{\rm bol}$ of the clumps are taken the values listed in the Table A1 of \citet{Liu2020}.
For the clumps as hub-filament systems, they should be in the state of global gravitational collapse, thus we assume  
$R_{\rm c}$=$r_{\rm in}$. 

For the inflow rate $\dot M$ of the clump, the tables in \citet{He2015-450,He2016-461} listed mass infall rates of 231 clumps. Based on the tables, we fitted the quantitative relation between the inflow rate and the mass of the clump, as shown in Fig.\ref{fit}.
Then, the infall rates of early-sources can be estimated using this relation according to their masses.
Finally, the infall velocity can be estimated by
\begin{equation}
v_{\rm in} =\frac{\dot M \times R_{\rm c}}{3 M_{\rm c}}. 
\end{equation}

\subsubsection{Radiation pressure}
Following \citet{Lopez2014-795},
the volume-averaged radiation pressure $P_{\rm rad}$ is 
\begin{equation}
P_{\rm rad} = \frac{3 L_{\rm bol}}{4 \pi R_{\rm c}^{2} c}.
\label{Prad}
\end{equation}
$L_{\rm bol}$ is the bolometric luminosity of the clump.

\subsubsection{Ionized gas pressure}
Following \citet{Lopez2014-795},
the ionized gas pressure is given by the ideal gas law, $P_{\rm HII} = (n_{\rm e} + n_{\rm H} + n_{\rm He}) kT_{\rm HII}$, where $n_{{\rm e}}$, $n_{\rm H}$, and $n_{\rm He}$ are the electron, hydrogen, and helium number densities, respectively, and $T_{\rm HII}$ is the temperature of the H{\sc ii} gas, which is assumed to be the same for electrons and ions. If helium is singly ionized, then $n_{{\rm e}} + n_{\rm H} + n_{\rm He} \approx 2 n_{\rm e}$. The electron temperature and density of resolved UC H{\sc ii} regions identified in the ATOMS survey have been derived from radio recombination lines H$_{\rm 40} \alpha$ and 3 mm continuum emission in \citet{Zhang2023-520}.
The corresponding velocities contributed by $P_{\rm rad}$ and $P_{\rm HII}$ are estimated by
$P_{\rm rad}= \rho v_{\rm rad}^{2}$ and $P_{\rm HII}= \rho v_{\rm HII}^{2}$.

\subsubsection{Relative importance of feedback processes}
From Table.\ref{early} and Table.\ref{late}, 
$v_{\rm HII}$ is significantly larger than other velocities, consistent with the conclusions in \citet{Lopez2014-795} that the ionized gas feedback dominates over the other considered feedback mechanisms.
In this work, we do not intend to study which feedback mechanism is the strongest, thus we ignored some feedback mechanisms, such as dust reprocessed radiation and stellar wind, 
which may also play important roles \citep{Pabst2019-565,Olivier2021-908,Henshaw2022-509}.
Here we note that the estimated $v_{\rm HII}$ is obviously larger than the observed line-widths in H$^{13}$CO$^+$ J=1-0, this is not surprising because the energy is deposited into the ionized gas and photodissociation regions. As expected,
late-sources present significantly stronger radiation pressure than early-sources. 
For early-sources, the velocity dispersion caused by the outflows is significantly smaller than that from the infalls and the radiation pressure.

\section{Discussion}

\subsection{Gravitational collapse under feedback}
\begin{figure*}
\centering
\includegraphics[width=0.95\textwidth]{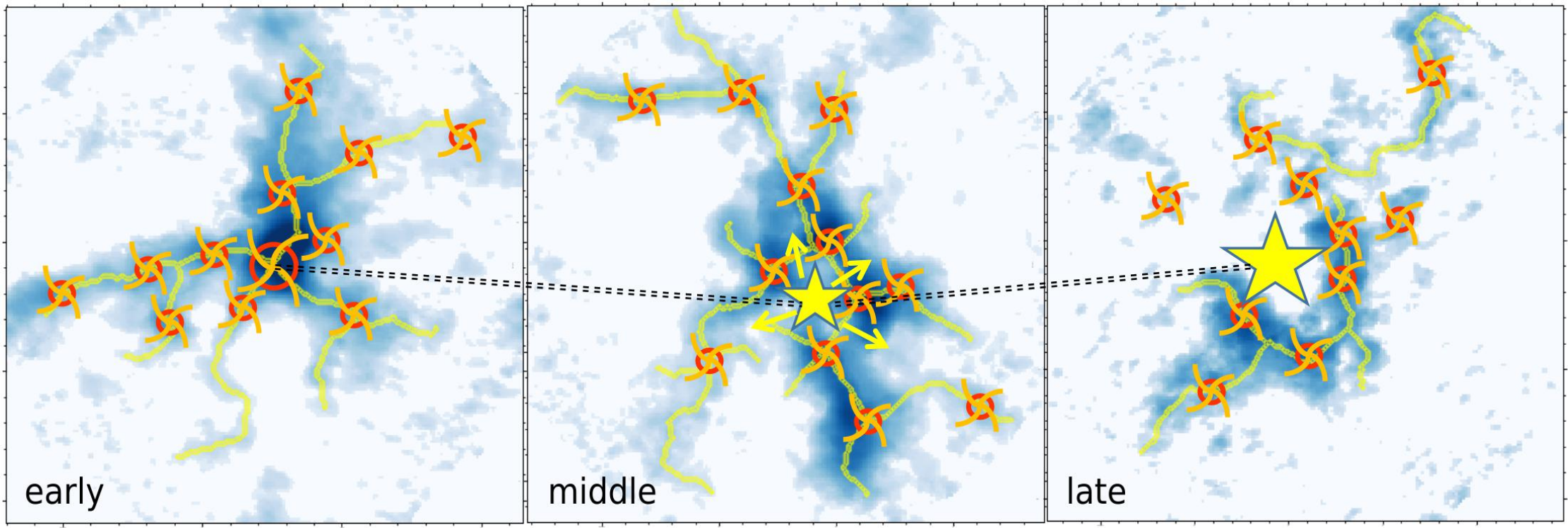}
\caption{Schematic diagram of the scenario argued in this work.
Evolution of a cloud with multi-scale hub-filament structures with increasing effects of feedback. Here we borrowed three pictures from Fig.\ref{map} as the background, for demo only.
Yellow lines are the filament skeletons identified in \citet{Zhou2022-514}.
Considering the competitive accretion scenario, the precursor of the protocluster is located in the central position of the hub. 
With the evolution of the protocluster, feedback from HII regions destroys the hub by mechanical force, i.e. the expanding HII regions push away and reshape the hub region, accompanied by the redistribution of small-scale dense structures which may be local hub-filament structures.
We note that small-scale hub-filament structures were not directly identified in this work due to the limitation of the resolution.}
\label{model}
\end{figure*}

As discussed in Sec.\ref{gradient-s}, even under strong feedback, the embedded dense structures in late-sources can still retain gravitational collapse. 
Feedback may redistribute local dense
gas structures, as illustrated in Fig.\ref{model}, because feedback does not seem to affect the
internal dynamics of the embedded dense gas structures.
The results are consistent with the conclusion in \citet{Zhou2023arXiv}:
Although the feedback disrupting the molecular clouds will break up the original cloud complex, the substructures of the original complex can be reorganized into new gravitationally governed configurations around new gravitational centers. This process is accompanied by structural destruction and generation, and changes in gravitational centers, but gravitational collapse is always ongoing.

Moreover, in
Fig.\ref{gradient}, we can find the same gas kinematic mode presented in \citet{Zhou2023-676}.
The variations of velocity gradients at small and large scales ($\sim$ 0.2\,pc as the boundary) are consistent with gravitational free-fall with central masses of $\sim$ 50\,M$_\odot$, $\sim$ 500\,M$_\odot$. It means that the velocity gradients on larger scales require larger mass to maintain, larger masses imply larger scales, that is to say, the larger scale inflow is driven by the larger scale structure which may be the gravitational clustering of smaller scale structures, consistent with the hierarchical/multi-scale hub-filament structures in the clouds and the gas inflow from large to small scales. 
In \citet{Zhou2023-676}, we studied the gas motions of the G333 molecular cloud complex, the longest filament is $\sim$50 pc. However, the largest scale in
Fig.\ref{gradient} is only $\sim$1 pc. The similar gas kinematic modes 
may indicate the self-similar kinematic structures on cloud-clump and clump-core scales, as illustrated in Fig.9 of \citet{Zhou2023-676}.

\subsection{The role of feedback}\label{feedback1}
In Sec.\ref{feedback},
the significantly larger v$_{\rm HII}$ will make us naturally think that late-sources should have the strongest feedback effects. However, that is not the case, as shown in Sec.\ref{basic}.
In Sec.\ref{basic}, overall, the velocity dispersion and virial ratios of the structures inside early- and late-sources are comparable. 
As shown in Sec.\ref{gradient-s}, the gas structures in late-sources are also in gravitational collapse like that in early-sources. Gravitational collapse may also have a significant contribution to the velocity dispersion of the gas structures in late-sources. Thus, the proportion of the velocity dispersion of the gas structures  contributed by expanding H{\sc ii} regions in late-sources decreases further.
The clouds exist as gas reservoirs, which contain dense gas structures as local gravitational centers.
Feedback from H{\sc ii} regions destroys the clouds by mechanical force, i.e. the expanding H{\sc ii} regions push away and reshape the loose cloud complex, some examples can refer to \citet{McLeod2015-450,Beuther2022-659,Bonne2023arXiv}. But the processes do not significantly affect the physical properties of the embedded dense gas structures in the clouds, suggesting that H{\sc ii} regions may not effectively inject momentum into the embedded dense gas structures as star-forming sites. 
However, the triggering does not necessarily result from momentum being transferred inside the gas structures but by compressing them, so they become gravitationally unstable.

Although feedback from H{\sc ii} regions is very strong, it is acting from the outside on the embedded dense gas structures. By comparison, the outflows and inflows can go deep inside of the embedded dense gas structures, and thus, their effect on the physical properties of the gas structures is not necessarily weaker than feedback from H{\sc ii} regions. We should not only focus on the strength of a feedback mechanism itself, but also assess the extent to which it can affect the embedded dense gas structures. 

In Fig.\ref{quantity}(e), the structures in middle-sources display slightly larger velocity dispersion than that in late-sources. Notably, middle-stage sources possess larger-scale hub-filament structures compared to their late-stage counterparts. The complex gas kinematics in these hub regions within middle-stage sources may contribute to the observed higher velocity dispersion. Furthermore, early-sources also exhibit large-scale hub-filament structures, but their gas structures present smaller velocity dispersion compared to those in middle-sources. This suggests that feedback from UC-H{\sc ii }regions in middle-sources plays a significant role. At least, we can conclude that feedback from protocluster-driven H{\sc ii} regions may not be the most powerful factor that affects the physical properties of local dense gas structures. It is important to note that the above discussion treats each clump in isolation, without considering the potential effects from the surrounding gas environment on the physical properties of the internal structures of the clumps, due to the limited FOV of ALMA observations. Given that the protocluster-driven H{\sc ii} regions may be still bound by the abundant surrounding gas, possible quasi-static expanding processes also indicate the limited effect of feedback from H{\sc ii} regions on the embedded dense gas structures. As shown in \citet{Kuhn2019-870}, star clusters and associations that are still embedded in molecular clouds are less likely to be expanding than those that are partially or fully revealed.

As illustrated in Fig.\ref{model},
hierarchical hub-filament structures of star-forming regions mean that star-forming regions exist as loose gravitational complexes. 
Feedback will reshape the overall structures of star-forming regions. 
This process is accompanied by the redistribution of the internal structures of the clouds, while the gravitational coupling between the internal structures is also reorganizing them into new gravitationally governed configurations around new gravitational centers. Generally, feedback will ultimately disperse the clouds and the reorganization is simply a resistance to the dispersal process.
The negative effects from feedback are that feedback destroys the gas reservoir inhabited by dense gas structures, inhibiting the further growth of dense structures by accreting the surrounding gas, thus suppressing the formation of massive stars. 
But oppositely, if feedback makes the distribution of dense gas structures more separated, and if these dense structures are still in a rich gas environment, the more separated distribution will improve the material accumulation efficiency of the dense structures by weakening the competitive accretion between them, and may eventually form more massive stars.

\section{Summary}

A total of 64 ATOMS sources at different evolutionary stages were selected to investigate the kinematics and dynamics of gas structures under feedback. 

1. We identified dense gas structures based on the 2D integrated intensity (Moment 0) map of H$^{13}$CO$^{+}$ (1$-$0) emission, and then extracted the average spectrum of each structure to investigate their velocity components and gas kinematics. We discussed the differences in the physical properties of different structure types at different evolutionary stages:
(1) Type3 structures with blended velocity components (refer to the Fig.6 of \citet{Zhou2023arXiv}) have larger velocity dispersion, higher column density and lower virial ratios than type1 structures with single velocity component at different evolutionary stages, which can be attributed to type3 structures being mainly located in hub regions.
(2) Overall, the velocity dispersion and virial ratios of the structures inside early- and late-sources are comparable. The structures inside middle-sources even have slightly larger velocity dispersion than that inside late-sources. 
These unusual results led us to question the effectiveness of feedback from H{\sc ii} regions in late-sources.

2. $\sigma-N*R$ always has stronger correlation compared to $\sigma-N$ and $\sigma-R$. 
There are significant correlations between velocity dispersion and column density, which may imply that the velocity dispersion originates from the gravitational collapse, also revealed by the measured velocity gradients.
Type1 and type3 structures present similar scaling relations, which further confirms that type3 structures are independent structures, not the superposition of uncorrelated foreground or background velocity components.

3. Dense gas structures in late-sources are in gravitational collapse like that in early-sources. Late-sources do not have large-scale hub-filament structures,  but the embedded dense gas structures in late-sources show similar kinematic modes to those in early- and middle-sources. These results may be explained by the multi-scale hub-filament structures in the clouds.

4. The differences in gravitational states of dense gas structures in different feedback environments were investigated by comparing local velocity gradients. The measured velocity gradients for the structures in early-sources and late-sources are comparable, indicating that feedback has a limited impact on the kinematic properties of local dense gas structures. The velocity gradients on larger scales require larger mass to maintain, larger masses imply larger scales, that is to say, the larger scale inflow is driven by the larger scale structure which may be the gravitational clustering of smaller scale structures, consistent with the hierarchical/multi-scale hub-filament structures in the clouds and the gas inflow from large to small scales. 

5. We quantitatively estimated the velocity dispersion generated by the outflows, inflows, ionized gas pressure and radiation pressure, and found that the ionized gas feedback is stronger than other feedback mechanisms. Late-sources present significantly stronger radiation pressure than early-sources.  For early-sources, the velocity dispersion caused by the outflows is significantly smaller than that from the infall and the radiation pressure. However, although feedback from H{\sc ii} regions is the strongest, it does not significantly affect the physical properties of the embedded dense gas structures in the clumps, suggesting that H{\sc ii} regions may not effectively inject momentum into the embedded dense gas structures as star-forming sites. 

6. This work on clump-core scales reaches the same conclusion with the study of the G333 molecular cloud complex on cloud-clump scales in \citet{Zhou2023arXiv}. Thus, we suggest that although
feedback from cloud to core scales will break up the original cloud complex, the substructures of the original complex can be reorganized into new gravitationally governed configuration around new gravitational centers. This process is accompanied by structural destruction and generation, and changes in gravitational centers, but gravitational collapse is always ongoing.

\begin{acknowledgements}
We would like to thank the referee for the detailed and constructive comments and suggestions that significantly improve and clarify this work. We thank J. H. He to share the tables of their studied infall samples.
This paper makes use of the following ALMA data: ADS/JAO.ALMA$\sharp$2019.1.00685.S. ALMA is a partnership of ESO (representing its member states), NSF (USA) and NINS (Japan), together with NRC (Canada), MOST and ASIAA (Taiwan), and KASI (Republic of Korea), in cooperation with the Republic of Chile. The Joint ALMA Observatory is operated by ESO, AUI/NRAO and NAOJ.

This work has been supported by the National Key R$\&$D Program of China (No. 2022YFA1603101). Tie Liu acknowledges the supports by National Natural Science Foundation of China (NSFC) through grants No.12073061 and No.12122307, the international partnership program of Chinese Academy of Sciences through grant No.114231KYSB20200009, and Shanghai Pujiang Program 20PJ1415500. PS was partially supported by a Grant-in-Aid for Scientific Research (KAKENHI Number JP22H01271 and JP23H01221) of JSPS. LB gratefully acknowledges support by ANID-BASAL project FB210003. MJ acknowledges support from the Research Council of Finland grant 348342.

\end{acknowledgements}

\bibliographystyle{aa} 
\bibliography{fed}


\appendix

\section{LTE analysis}\label{lte-e}

Following the procedures described in \citet{Garden1991-374, Mangum2015-127}, for a rotational transition from upper level J + 1 to lower level J, we can derive the total column density by
\begin{equation}
 N_{tot} =\frac{3k}{8\pi^3{\mu}^2B(J+1)}\frac{(T_{\rm ex}+hB/3k)\exp[hBJ(J+1)/kT_{\rm ex})]}{1-\exp(-h\nu/kT_{\rm ex})}
 \int{\tau{\rm dv}},
\label{}
\end{equation}
\begin{equation}
\tau ={\rm-ln}[1-\frac{T_{\rm mb}}{J(T_{\rm ex})-J(T_{\rm bg})}],
\end{equation}

\begin{equation}
\int{\tau{\rm dv}} = \frac{1}{J(T_{ex}) - J(T_{bg})} \frac{\tau}{1-e^{-\tau}}
\int{T_{\rm mb}{\rm dv}}
\label{tem}
\end{equation}

\begin{equation}
J(T) = \frac{h\nu/k}{e^{h\nu/kT}-1}
\label{j}
\end{equation}
where $B=\nu/[2(J+1)]$ is the rotational constant of the molecule, $\rm \mu$ is the permanent dipole moment, and $\rm \mu= 3.89$ Debye for H$^{13}$CO$^{+}$. $\rm T_{bg}=2.73$ K is the background temperature, $\int{T_{\rm mb}{\rm dv}}$ represents the integrated intensity. In the above formulas, the correction for high optical depth was applied \citep{Frerking1982-262,Goldsmith2008-680,Areal2019-36}.
Assuming optically-thick emission of HCO$^{+}$ (1-0) emission, 
we can estimate the excitation temperature $T_{\rm ex}$ following the formula \citep{Garden1991-374,Pineda2008-679}
\begin{equation}
T_{\rm ex} = \frac{4.28}{{\rm ln}[1 + 1/(0.23T_{\rm peak}+0.26)]}, 
\label{tex}
\end{equation}
where T$_{\rm peak}$ is the observed HCO$^{+}$ (1-0) peak brightness temperature.

\section{The maps of all samples}\label{all}
\begin{figure*}
\centering
\includegraphics[width=0.95\textwidth]{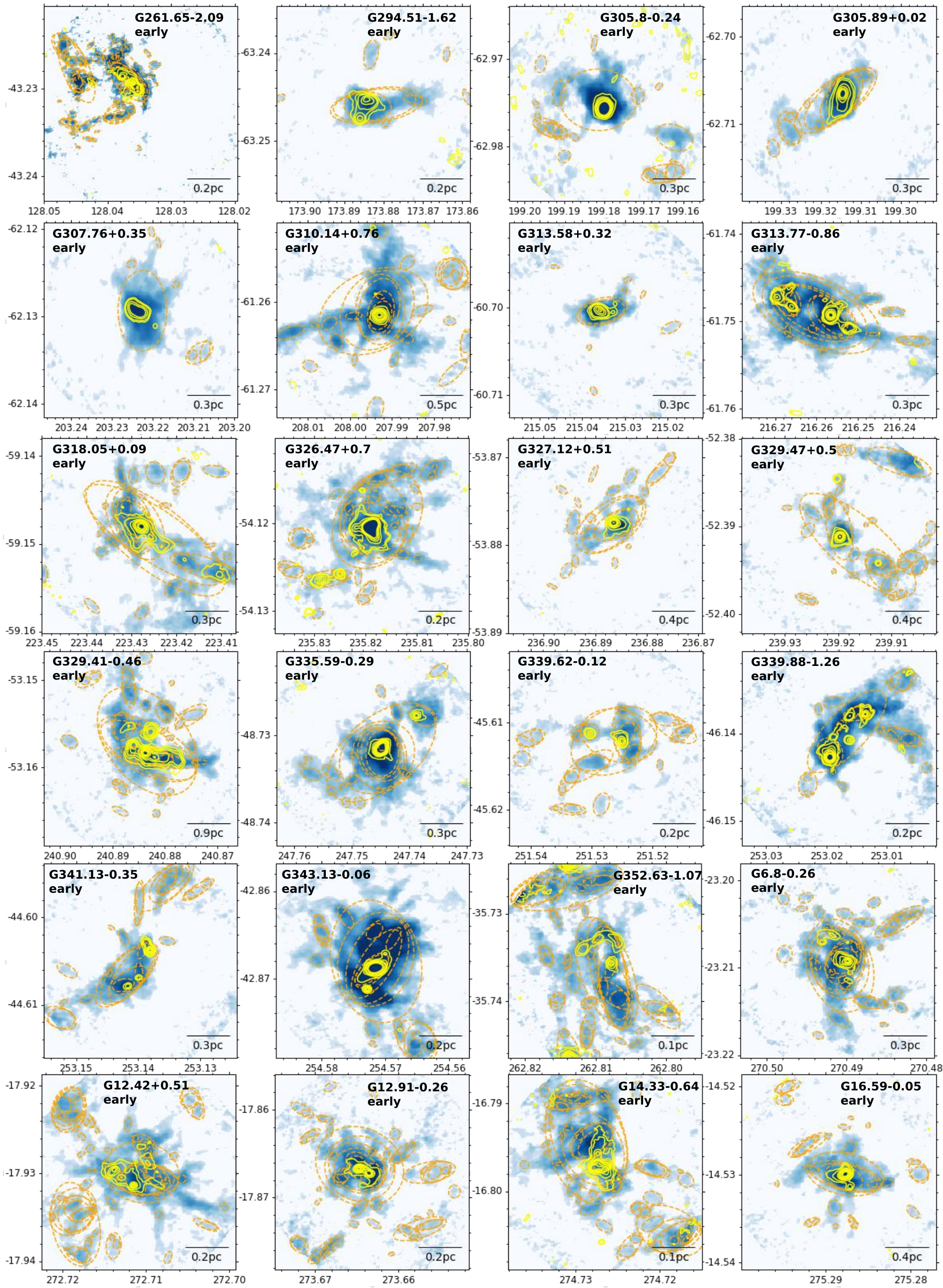}
\caption{Same as Fig.\ref{map}.}
\label{map1}
\end{figure*}

\begin{figure*}
\centering
\includegraphics[width=0.95\textwidth]{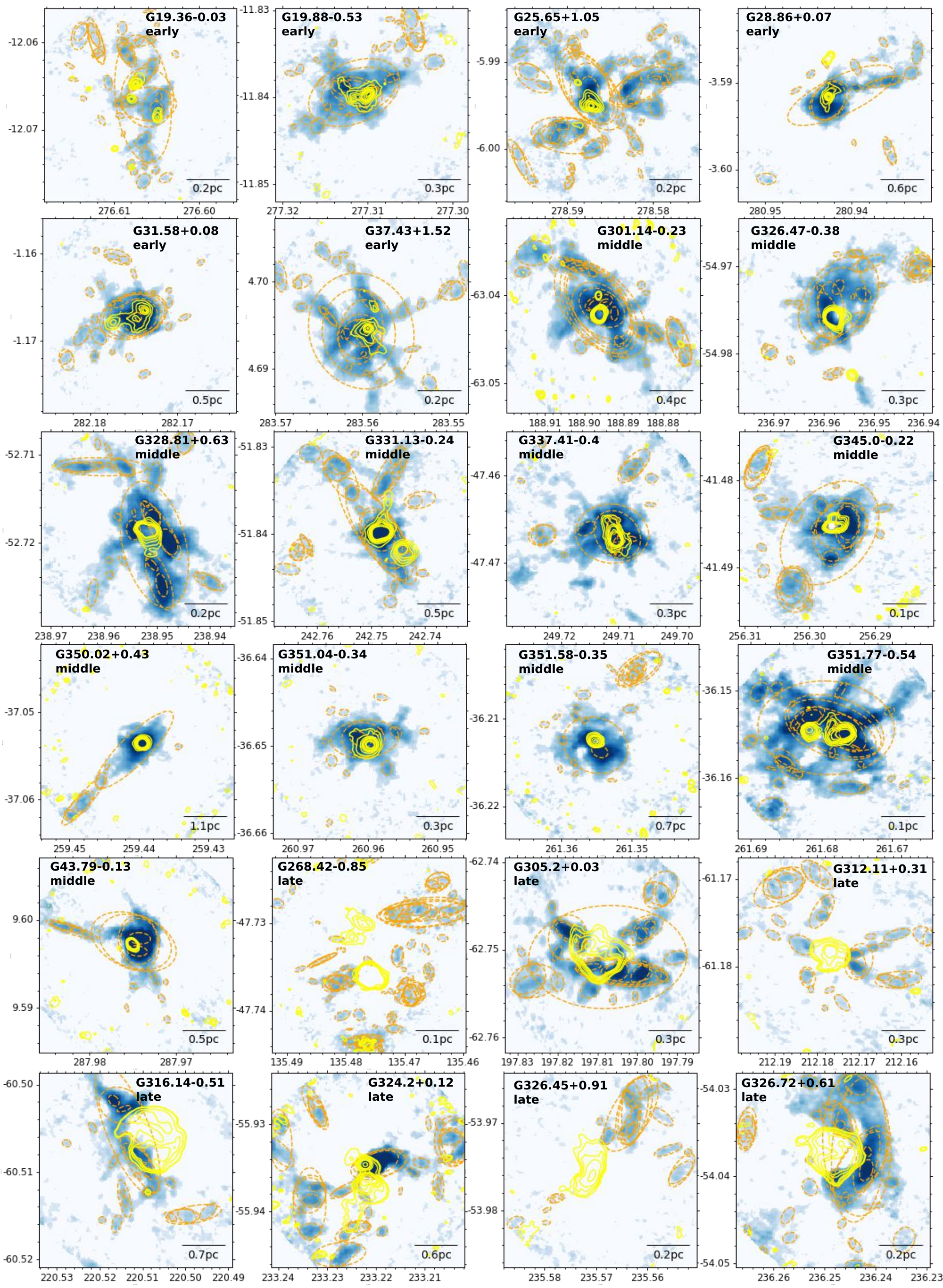}
\caption{Same as Fig.\ref{map}.}
\label{map2}
\end{figure*}

\begin{figure*}
\centering
\includegraphics[width=0.95\textwidth]{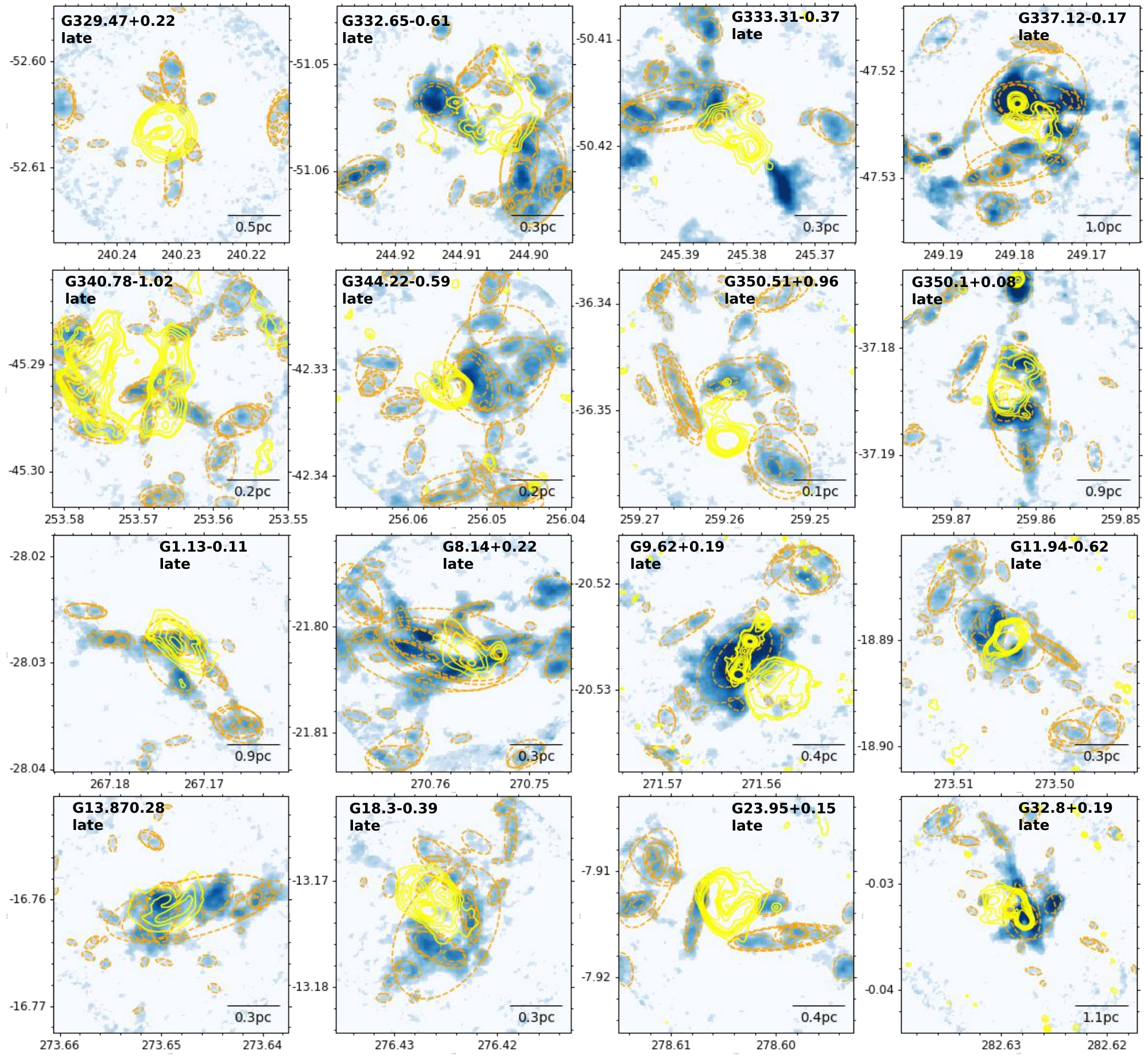}
\caption{Same as Fig.\ref{map}.}
\label{map3}
\end{figure*}

\end{document}